\documentclass[fleqn,usenatbib]{mnras}

\usepackage{newtxtext,newtxmath}

\usepackage[T1]{fontenc}

\DeclareRobustCommand{\VAN}[3]{#2}
\let\VANthebibliography\thebibliography
\def\thebibliography{\DeclareRobustCommand{\VAN}[3]{##3}\VANthebibliography}

\usepackage{graphicx}	
\usepackage{amsmath}	
\usepackage{courier}    
\title[Identification of Cosmic Chronometers in the GAMA Survey]{Identification of Cosmic Chronometers in the GAMA Survey}

\author[L. J. Hunt et al.]{Laura J. Hunt,$^{1,2,3}$
Kevin A. Pimbblet,$^{2,3}$
and David M. Benoit$^{2}$
\\
$^{1}$The Open University, Walton Hall, Kents Hill, Milton Keynes, MK7 6AA, UK \\
$^{2}$E.A. Milne Centre for Astrophysics, Faculty of Science and Engineering, University of Hull, Cottingham Road, Kingston-upon-Hull HU6 7RX, UK \\
$^{3}$ Centre of Excellence for Data Science, AI, and Modelling (DAIM), University of Hull, Cottingham Road, Kingston-upon-Hull, HU6 7RX, UK
}

\date{Accepted XXX. Received YYY; in original form ZZZ}

\pubyear{2026}

\begin{document}
\label{firstpage}
\pagerange{\pageref{firstpage}--\pageref{lastpage}}
\maketitle

\begin{abstract}
We present a pure sample of 231 massive, passively evolving elliptical galaxies as cosmic chronometer candidates selected from the Galaxy and Mass Assembly (GAMA) survey between $0.03<z<0.4$. The goal of this work is to increase the pool of high-purity cosmic chronometer candidates available in the literature and assess the effect of purity on statistical uncertainty to make $H(z)$ and $H_0$ calculations more robust in future studies. We use traditional selection criteria to identify cosmic chronometers using photometry and spectroscopy in conjunction with modelled parameters such as stellar mass and velocity dispersion. We use an NUVrJ photometric selection to identify quiescent galaxies and identify a number of spectral features, including H$\alpha$, [OII] and H$\delta$ to use as tracers for ongoing or recently halted star formation. Additionally, as part of the selection, we calculate the ratio between the Ca II H and K lines to determine whether a galaxy's stellar population is dominated by older or younger stars. We confirm CC candidacy by selecting only the most massive galaxies and perform a visual inspection of the spectra and images of CC candidates. We provide an analysis of the parent sample in comparison to the CC sample and confirm CCs have much lower sSFR, with a median sSFR of $10^{-12}~\text{Myr}^{-1}$. Additionally, we find their median ages are $\sim2~\text{Gyr}$ older than the parent sample. 

\end{abstract}




\section{Introduction}
The nature of universal expansion remains one of the most important unanswered questions in modern cosmology especially when considering that there is tension in the agreed upon measurement of the expansion rate, the Hubble constant $H_0$ \citep[for reviews:][]{2021CQGra..38o3001D, 2021ApJ...919...16F, 2021A&ARv..29....9S}. For example, recent early-Universe measurements of $H_0$ found using the Cosmic Microwave Background (CMB) temperature and polarisation anisotropies and the $\Lambda$CDM model find $H_0=67.4\pm0.5$~kms$^{-1}$~Mpc$^{-1}$ \citep{2020A&A...641A...6P} whereas late-Universe distance ladder calibrations of Cepheids and type Ia supernovae find $H_0=73.2\pm1.3$~kms$^{-1}$~Mpc$^{-1}$ \citep{2022ApJ...934L...7R} which leaves us with tensions between $\sim4-6\sigma$. 

Typically, we characterise the disparity between results based on different calculations as stemming from issues relating to either model assumptions or incorrect physics from $\Lambda$CDM models used for early-Universe measurements of $H_0$ or the systematic uncertainties that compound when calibrating the distance ladder for late-Universe measurements. It is therefore obvious that alternative methods must be used to reconcile the problem and determine whether our physics needs revision or our measurements need to be more accurate. One such method involves directly probing the expansion rate in the late-Universe by using cosmic chronometry \citep{2002ApJ...573...37J}.

This method describes the rate of cosmic expansion at any given time using the Hubble parameter $H(z)$ which is expressed in terms of the scale factor $a(t)$ and the derivative of the scale factor $\dot a(t)$. The scale factor is related to redshift by $a=(1+z)^{-1}$ such that:

\begin{equation}
    H(z)\equiv \frac{\dot{a}}{a} \equiv \frac{-1}{1+z}\frac{\Delta z}{\Delta t}
\end{equation}

where $\Delta t$ and $\Delta z$ quantify the difference in age and redshift between two objects with an effective redshift $z$ which allows us to trace the rate at which redshift changes with cosmic time. By using this method we can constrain the expansion history of the Universe, independently from cosmology through direct measurements of expansion itself \citep[e.g.][]{2014PDU.....5..307V, 2019JCAP...03..043J, 2019MNRAS.483.4803L, 2022JHEAp..36...27V, 2022ApJ...928L...4B, 2023ApJS..265...48J, 2025MNRAS.540.3135L, 2025MNRAS.544.3064L, 2026A&A...707A.111T, 2025arXiv251202109T, 2025A&A...696A..98T}. However, this relationship relies on the idea of ``standard clocks'' or cosmic chronometers as they have a uniform formation history that allows them to be used to measure the age of the Universe. Various objects may be used as CC candidates \citep[e.g.][]{2022JHEAp..36...27V, 2025MNRAS.540.3135L, 2026A&A...707A.111T, 2025arXiv251202109T, 2025A&A...696A..98T} however one of the most reliable and well studied candidates are massive, passively evolving elliptical galaxies as they have simple, well-defined star formation histories (SFHs) with extremely old stellar populations that allow us to trace cosmic time across redshift. 

Ideal cosmic chronometer (CC) candidates are extremely homogeneous in both evolution and redshift, therefore they typically form their stellar populations at high redshift $z>2$ with less than 1\% of the stellar mass forming after $z<1$ \citep{2014PDU.....5..307V, 2016JCAP...12..039M}. By taking these passive galaxies at different redshifts, we can compare the upper cutoff of their age distributions in order to produce a more statistically significant measurement of $\Delta z/\Delta t$ and therefore $H(z)$. \cite{2005PhRvD..71l3001S} and \cite{2010JCAP...02..008S} use the differential age-redshift relationship to calculate $H(z)$ measurements for pairs of passive galaxies between $0.2<z<1.0$. \cite{2010JCAP...02..008S} notes the importance of selecting galaxies with minimal contamination from younger stellar populations to avoid causing systematically younger ages. \cite{2018JCAP...10..015H} find that combining low-$z$ $H(z)$ measurements derived from two CC regimes with baryonic acoustic oscillations and supernovae measurements reduces the scatter in $H_0$. They report tensions of 1.41$\sigma$ and 2.60$\sigma$ when comparing their results to \cite{2016A&A...596A.107P} and \cite{2018ApJ...855..136R} respectively. \cite{2021ApJ...908...84V} finds similar results when using $H(z)$ values, derived from CC studies, to constrain spatial curvature. From this, they argue that CCs are extremely important to solving the Hubble tension as they are free from cosmological model assumptions.

Massive, passively evolving galaxies are ideal candidates for cosmic chronometers due to the nature of their rapid formation as their stellar populations are incredibly uniform and they do not exhibit contamination from a younger stellar population that would introduce uncertainty in their age determination. However, objects such as these, with very consistent evolution histories and complete passivity are extremely rare as most galaxies experience at least some minor mergers over the course of their evolution which can reignite star formation. Typically, studies can go down two routes to minimise the uncertainty in $H_0$ by either maximising the total number of potential chronometers by using less strict selection criteria \citep[e.g.][]{2016JCAP...05..014M, 2025MNRAS.544.3064L} but risk systematic uncertainties associated with mixed stellar populations. Or use much stricter selection criteria to create a purer sample of chronometers to trace expansion more tightly. The downside with the later option is that using stricter selection criteria produces much smaller samples of potential CC candidates which are typically limited to a few hundred galaxies \citep[e.g.][]{2022ApJ...928L...4B, 2023ApJS..265...48J, 2023A&A...679A..96T} which is why it is vital to produce more studies that utilise different surveys to increase this sample. Therefore, a fundamental part of this work relies on the selection of good cosmic chronometer (CC) candidates that contain little or no contamination from younger stellar populations with the aim of finding a high-purity CC sample in the GAMA survey. 

The criteria we use are based on the traditional selection criteria that \cite{2018ApJ...868...84M} describes. They quantify the impact of contamination from younger stellar populations on the overall population and develop a selection pipeline designed to minimise this contamination as much as possible. Typically, studies in this field use a photometry-based selection using UVJ or NUVrJ diagrams to select for quiescent galaxies \citep[e.g.][]{2023A&A...679A..96T, 2023ApJS..265...48J}. Additionally, some combination of spectral features such as H$\alpha$ and [OII] emission lines as well as H$\delta$ absorption lines and the H:K ratio are used as these are tracers for ongoing and recently halted star formation \citep[e.g.][]{2011JCAP...03..045M, 2022ApJ...928L...4B, 2025MNRAS.544.3064L}. Finally, a selection based on stellar mass or stellar velocity dispersion is taken in order to select the oldest, most massive galaxies with the most synchronised SFHs before a final visual inspection is taken and only then can an object be considered truly a good CC candidate. In this work, we will identify cosmic chronometer candidates using traditional selection criteria and assess their properties. In a follow-up to this paper, we will use this pure sample of CCs to train a machine learning model and assess its capabilities at CC detection in order to create large samples of CCs to calculate $H(z)$ and $H_0$.

Firstly, we describe the selection of cosmic chronometers from the Galaxy and Mass Assembly (GAMA) survey \citep{2009A&G....50e..12D, 2015MNRAS.452.2087L, 2018MNRAS.474.3875B} in section~\ref{data}. This includes an evaluation of the different selection criteria, the results of which we discuss in section~\ref{analysis}. To evaluate the significance of a purer CC sample, we calculate and compare the $H(z)$ and its statistical uncertainties of different subsamples of CCs in section~\ref{stat_unc}. Finally, we compare our selection of cosmic chronometers in the GAMA survey to previous work in section~\ref{disc}. Some assumed cosmology is required for stellar population synthesis done by the GAMA survey and the SED fitting used in MagPhys, in which they use $H_0=70$~kms$^{-1}$~Mpc$^{-1}$, $\Omega_\text{m}=0.3$ and $\Omega_{\Lambda}=0.7$. Additionally, we use an assumed cosmology to estimate the corresponding cosmic time across redshift bins in the binning process in section~\ref{binning}, however, as this is only a loose estimate of cosmic time needed for approximate bin sizes we do not believe there to be a dependence on the assumed cosmology here.

\section{Data and Selection Criteria} \label{data}
\begin{figure*}
    \centering
    \includegraphics[width=\linewidth]{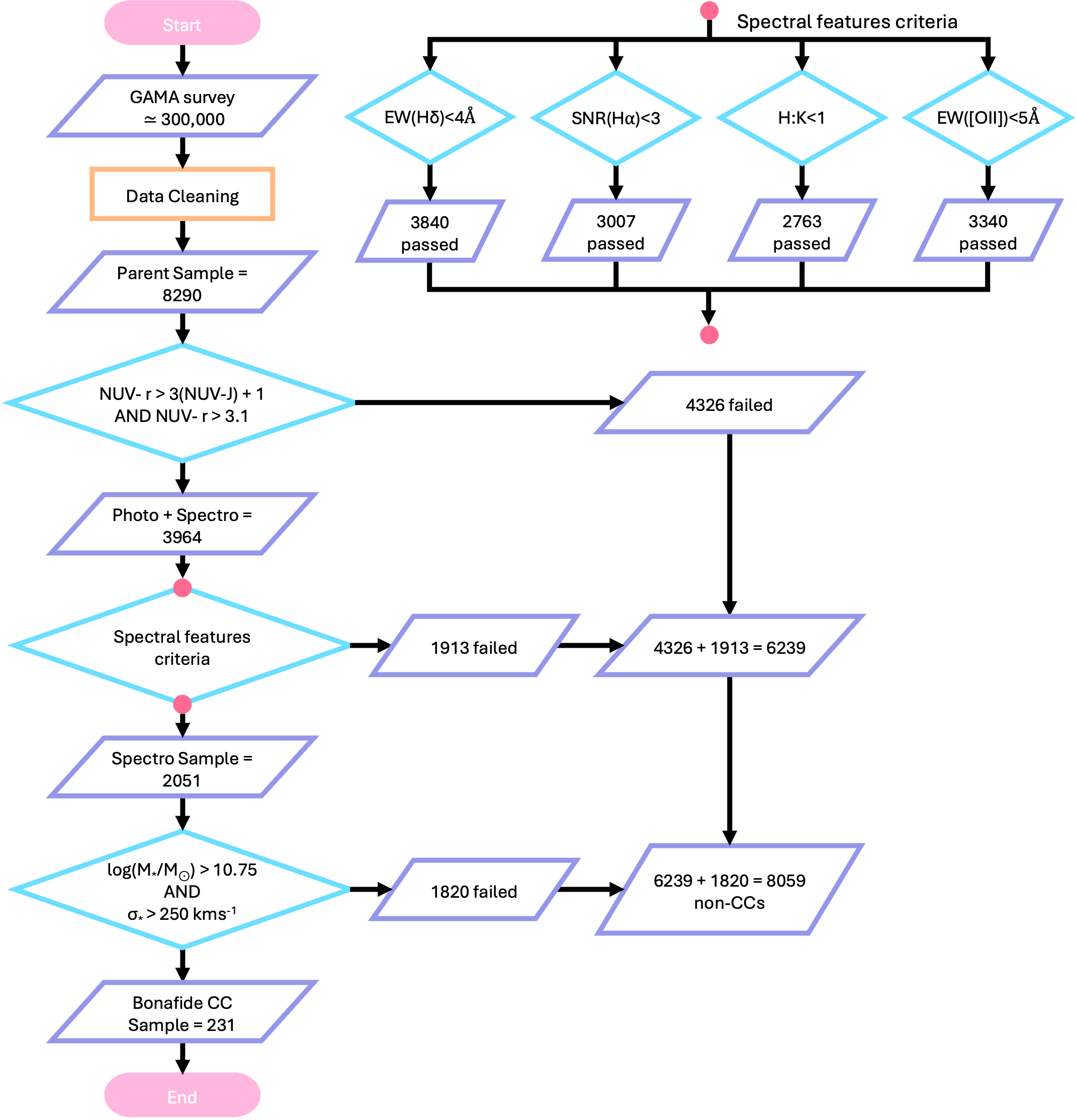}
    \caption{For clarity, this flowchart shows the process of selecting cosmic chronometers from the GAMA survey using traditional selection criteria. Each selection cut is made sequentially whereas the four spectroscopic cuts are made concurrently (ie. a galaxy must pass all four cuts to pass the spectroscopic cut). We show the number of galaxies that pass each of the four spectroscopic cuts separately along with the total number that pass all four tests. Additionally, for total clarity we note how many galaxies fail each test and the total number of CCs and non-CCs identified.}
    \label{fig:trad_cuts}
\end{figure*}

We use photometric and spectroscopic data collected by the Galaxy And Mass Assembly (GAMA) \citep{2009A&G....50e..12D, 2015MNRAS.452.2087L, 2018MNRAS.474.3875B} survey which observes $\sim300,000$ galaxies over $\sim286$~deg$^2$, down to a limiting magnitude of $r<19.8$~mag. The GAMA survey provides spectra using the AAOmega spectrograph on the Anglo-Australian Telescope and catalogue-matched photometry from various independent surveys including GALEX, VIKING and SDSS imaging data using \texttt{SExtractor}. Additionally, using high-quality optical spectra, GAMA derives stellar population parameters such as mass-weighted age, star formation rate and metallicity using the \texttt{MAGPHYS} SED-fitting code \citep{2008MNRAS.388.1595D} and stellar masses using stellar population synthesis modelling of broadband photometry. We show the selection process as an easy-to-read flowchart in Fig.~\ref{fig:trad_cuts} which shows the number of galaxies that pass and fail at each selection cut. We first complete basic data cleaning following the method described by GAMA which yields 8,290 galaxies with everything required for classification.

\subsection{Photometric Selection: UV Emission}
Following the \cite{2013A&A...556A..55I} prescription, we select galaxies with NUV~$- r > 3$~(NUV~$-J) + 1$ and NUV~$- r > 3.1$, as shown in Fig.~\ref{fig:nuvrj}, to minimise the contamination from young stellar populations with ages between $0.1-1$~Gyr. We use NUV magnitudes from the GalexMain table from the \texttt{GalexPhotometry v02} DMU \citep{2015MNRAS.452.2087L} provided by GAMA DR3. We use $r$ and $J$ band magnitudes from the ApMatchedCat table from the \texttt{ApMatchedPhotom v06} DMU \citep{2016MNRAS.455.3911D} provided by GAMA DR3. NUV, $r$ and $J$ magnitudes are k-corrected using corrections provided in the kcorr\_auto\_z00 table in the \texttt{kCorrections} \citep{2012MNRAS.420.1239L} and corrected for galactic extinction using those given in the GalacticExtinction table in the \texttt{EqInputCat v46} from GAMA DR3 \citep{2010MNRAS.404...86B}. By using this cut for UV emission, we exclude galaxies with stellar populations in the $0.1-1$~Gyr age range which results in a yield of 3,964 galaxies.

\subsection{Spectroscopic Selection:}
\subsubsection{H:K Ratio}
We calculate the ratio between the calcium II H and K lines using restframe GAMA spectra from the SpecObj table of the \texttt{SpecCat v27} DMU \citep{2015MNRAS.452.2087L}. We calculate the ratio between the Ca II H and K lines at $\lambda 3969$~{\AA} and $\lambda 3934$~{\AA} respectively following the method described by \cite{2019MNRAS.490..455P}. This method describes calculating the depth of troughs with respect to the continuum line between a zero-point range of $3948-3955$~{\AA} to ensure the line strength is significant in comparison to the continuum. The standard deviation of the pixel flux in the zero-point range is used to estimate a SN ratio which we then use to ensure the troughs are deep enough for the H:K ratio to be meaningful. Galaxies with a H:K ratio $< 1$ are dominated by older stellar populations so we take this to be our cut to exclude stellar populations $< 200$~Myr old \citep{2007MNRAS.381..543W, 2018ApJ...868...84M}.

\subsubsection{H$\delta$ Absorption}
We use H$\delta_A$ (hereafter, H$\delta$) EW from the GAMA survey EW measurements from the Direct Summation table of the \texttt{SpecLineSFR} DMU \citep{2017MNRAS.465.2671G}. We then take a cut for galaxies that have an absorption EW(H$\delta) < 4$~{\AA} in order to exclude post-starburst galaxies or galaxies with stellar populations that are $0.5-1$~Gyr old. 

\subsubsection{H$\alpha$ and [OII] Emission}
Similarly to H$\delta$ we take GAMA survey EW measurements from the Direct Summation table of the \texttt{SpecLineSFR} DMU \citep{2017MNRAS.465.2671G}. For [OII] $\lambda 3727$ emission, we make a cut for EW([OII])~$< 5$~\AA~to select only galaxies with no detectable emission \citep{2009A&A...493...39M}. We then select galaxies with emission SNR(H$\alpha) < 3$, we use SNR as opposed to a direct EW measurement because this wavelength region is typically well detected and suffers less from dust effects. Additionally, we care more about the presence of the line rather than its relative strength \citep{2018MNRAS.474.1873W}. These two cuts avoid galaxies with stellar populations younger than $\sim10$~Myr.

After these four spectroscopic cuts we yield 2,051 candidates.

\subsection{Stellar Mass, Velocity Dispersion and Visual Inspection}
For stellar mass, we use those provided in the StellarMasses v19 table in the \texttt{StellarMasses v20} DMU \citep{2011MNRAS.418.1587T} for the GAMA survey. GAMA stellar masses are derived from aperture-matched photometry (i.e. the $r-$~defined AUTO photometry from the \texttt{ApMatchedCat} table in the \texttt{ApMatchedPhotom} DMU \citep{2011MNRAS.412..765H, 2015MNRAS.452.2087L, 2015arXiv150700665D} however the AUTO photometry is not necessarily representative of the total flux in any particular band. Therefore, it is recommended to make an aperture correction to the stellar mass which is given as the fluxscale. Once corrected for fluxscale we take a cut for galaxies with log$(M_{*}/M_{\odot}$)~>~10.75. Stellar velocity dispersions are taken from SDSS DR17 \citep{2012AJ....144..144B} and cross-matched with our GAMA galaxies which we then take a cut for galaxies with $\sigma_{*}$~>~250~kms$^{-1}$. Combining these cuts selects for the oldest galaxies with the most synchronised SFHs.

After cutting lower-mass and $\sigma_*$ galaxies, we visually inspect the spectra and any available imaging of the CC candidates to ensure the final sample can be considered good CCs which leaves us with 231 CC candidates from the GAMA survey. It is important to visually inspect the images and spectra of CC candidates as even some of the selected galaxies may exhibit intermediary properties causing misclassification \citep{2009A&A...493...39M}. We find no evidence of misclassification and yield a high-purity sample of 231 cosmic chronometer candidates. 

\subsection{Age and Other Properties}
With our selected CC sample, we compare the stellar population properties of the parent sample and each of the subsequently selected samples. These properties are generated using the spectral energy distribution (SED) fitting program MagPhys \citep{2008MNRAS.388.1595D} and are provided by GAMA in the MagPhys table of the \texttt{MAGPHYS v06} DMU. MagPhys is a physically motivated code that interprets galaxy emission across ultraviolet, optical and infrared wavelength ranges using spectral templates of stellar populations generated from stellar population synthesis code by  \cite{2003MNRAS.344.1000B} in comparison with observed photometry whilst taking dust into account in order to identify the best overall fit stellar and dust template pair for which the associated stellar population properties are taken from. A probability density function (PDF) is calculated for each property with 2$\sigma$, 1$\sigma$ and median percentiles given.

MagPhys parameterises star formation histories in the stellar population library following \cite{2003MNRAS.341...33K} methods by characterising an underlying continuous model by an age $t_{g}$ and a star formation time-scale parameter $\gamma$ and introduces random bursts on the continuous model. In which the underlying continuous model has exponentially declining star formation rates, 

\begin{equation} \label{eq:sfr}
    \psi(t)\propto \exp{(-\gamma t)}
\end{equation}

where $\gamma$ is the star formation time-scale parameter which corresponds to models with $\gamma=  0, 0.07 \text{~and~} 0.25 \text{~Gyr}^{-1}$ at ages $t= 1.4, 10 \text{~and~} 10 \text{~Gyr}^{-1}$ which represent starburst, normal star-forming and quiescent star-forming galaxies. It is important to mention that biases may be introduced by their attempt to avoid oversampling galaxies with negligible current star formation and the inclusion of random bursts that occur with equal probability at any given time until $t_{g}$. In addition, they state that the likelihood of a galaxy having experienced a burst in the last 2~Gyr is set to 50\%. As such, we use the values for median mass-weighted age, specific SFR, SFR, metallicity and time since most recent starburst $t_{\text{starburst}}$ and time since formation $t_{\text{form}}$ that can be found in the \texttt{MAGPHYS} DMU provided by GAMA.

\section{Comparison of Selection Criteria and Properties} \label{analysis}
To evaluate our bona fide cosmic chronometers against the parent sample, we compare observations used for the selection criteria and modelled properties from the GAMA survey estimated with MagPhys SED fitting code. The four samples we created in section~\ref{data} are based on which galaxies pass each subsequent selection test and we define them as such:

\begin{enumerate}
    \item The ``parent'' sample comprises all 8290 galaxies
    \item The ``photo'' sample includes all 3964 galaxies that pass the NUVrJ prescription indicating quiescence
    \item The ``spectro-photo" sample includes all of the galaxies that pass the NUVrJ prescription and the spectroscopic criteria but fail the mass and velocity dispersion criteria
    \item The final ``bona fide'' sample only contains the galaxies that pass all criteria, including high stellar mass, velocity dispersion and visual inspection checks
\end{enumerate}

\begin{figure}
    \centering
    \includegraphics[width=\linewidth]{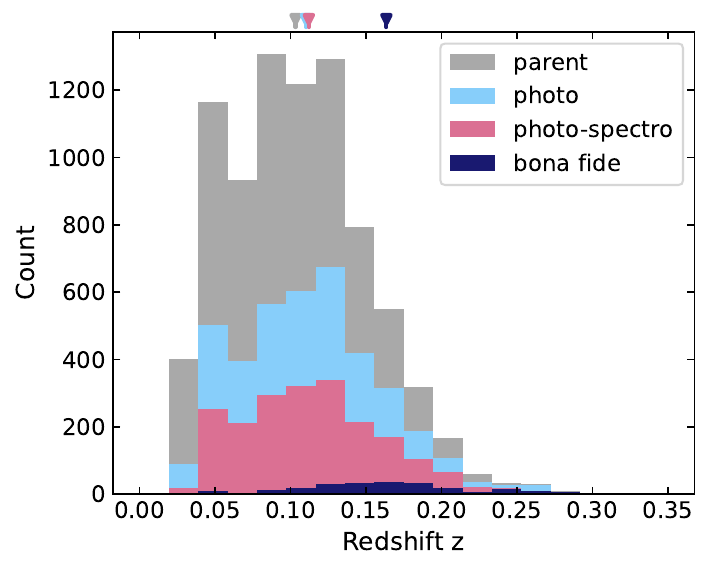}
    \caption[]{The distribution of redshift across the different samples. The parent sample is shown in grey whilst the photo, photo-spectro and bona fide samples are shown in blue, pink and purple respectively. The median value for each sample is shown with an arrow and reported in Table~\ref{tab:med_vals1}.}
    \label{fig:redshift}
\end{figure}

In Fig.~\ref{fig:redshift}, we show the distribution of redshifts for each of the four samples. The median redshift for the bona fide CC subsample is $\sim0.05$ higher than that of the parent, photo and photo-spectro subsamples. This may be a result of the selection bias due to the mass completeness of the parent sample, which we investigate in section~\ref{binning}. 

\begin{figure}
    \centering
    \includegraphics[width=\linewidth]{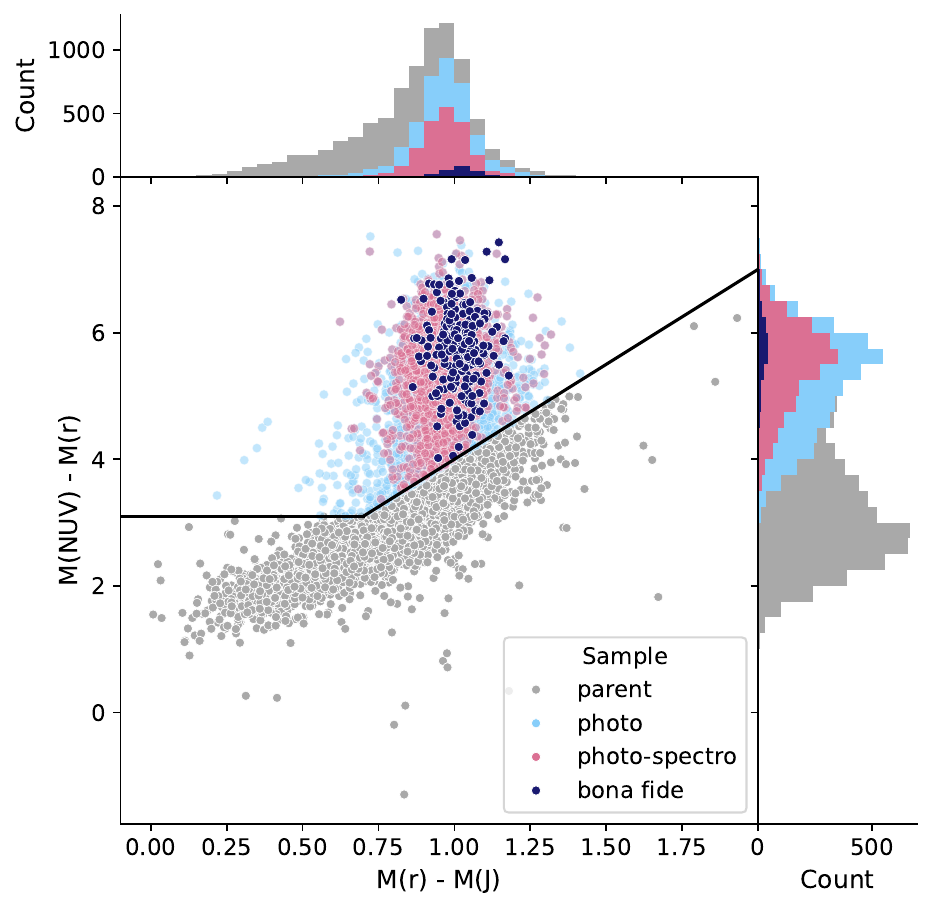}
    \caption{NUVrJ diagram showing the star-forming and quiescent populations from our dataset. Galaxies above the black line are classified as passives. Grey points below this prescription are classified as star-forming. The blue, pink and purple groups show the mix of passive galaxies in the photo, photo-spectro and bona fide samples respectively.}
    \label{fig:nuvrj}
\end{figure}

Fig.~\ref{fig:nuvrj} shows the distribution of each of the four samples in relation to their UV emission. Galaxies that sit above the black line, in which NUV~$- r > 3$~(NUV~$-J) + 1$ and NUV~$- r > 3.1$ are considered quiescent, whereas galaxies below this prescription are considered star-forming and removed from the subsequent three selection samples \citep{2013A&A...556A..55I}. The ``photo'' sample that pass this photometric selection cut but not the final two cuts for spectral features and stellar properties are shown in blue. The bimodality between redder, quiescent galaxies and bluer, star-forming galaxies is demonstrated in the histograms in Fig.~\ref{fig:nuvrj} as there are two clear peaks when considering NUV$-$ and $r-$band colours.

\begin{figure*}
    \centering
    \includegraphics[width=\linewidth]{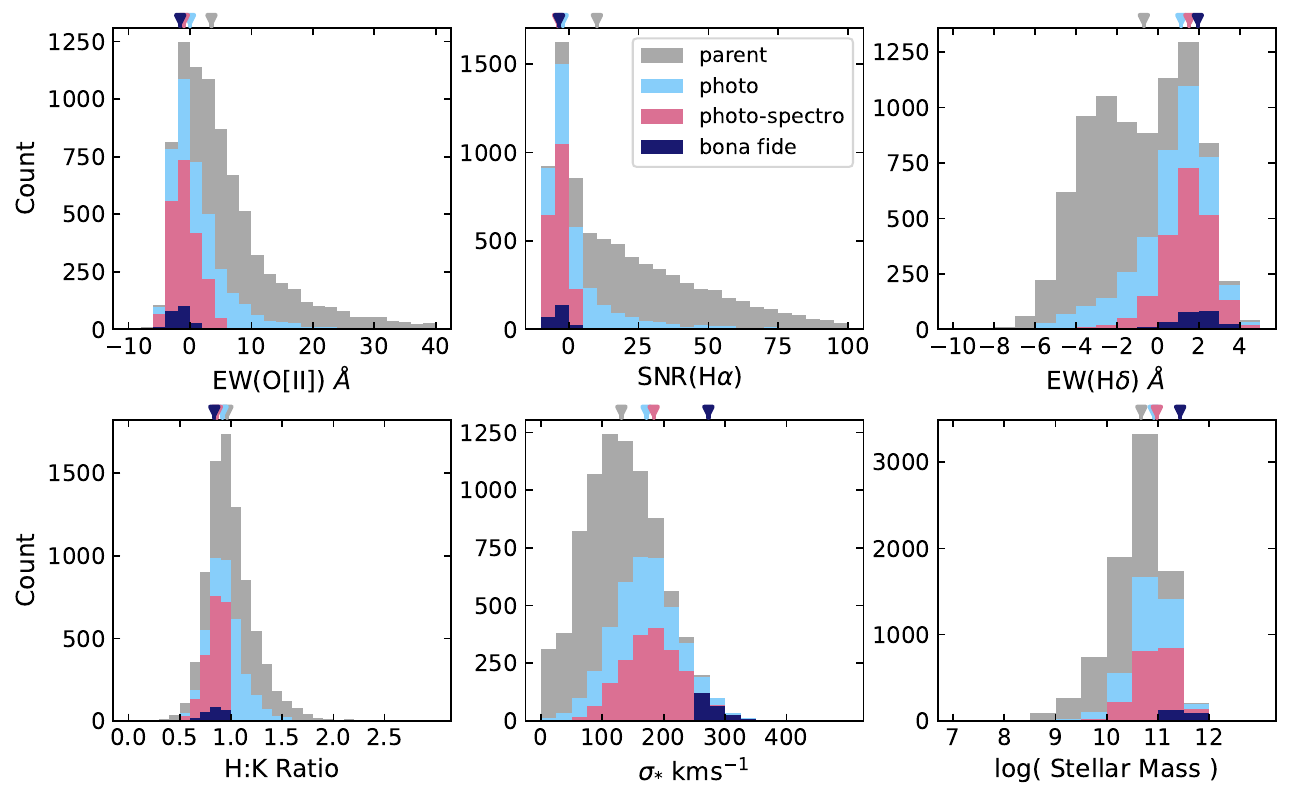}
    \caption[]{The distribution of each selection sample with each cut. The parent sample is shown in grey whilst the photo, photo-spectro and bona fide samples are shown in blue, pink and purple respectively. The upper panels from left to right show the EW([OII]) and SNR(H$\alpha$) emission cuts and the EW(H$\delta$) absorption cut. The lower panels from left to right show the distribution of the samples with the H:K ratio, stellar velocity dispersion and stellar mass respectively. The median value for each sample is shown with an arrow.}  
    \label{fig:mcuts}
\end{figure*}

\begin{table*} 
    \centering 
    \caption{The number of galaxies in each sample after each set of selection cuts. For each sample we report the median values and standard deviations of their redshift, H$\alpha$, [OII], H$\delta$ EWs, H:K ratio and stellar velocity dispersion.} 
    \label{tab:med_vals1} 
    \begin{tabular}{c|ccccccccccccc} 
    \hline 
    Sample & Count & $\langle z \rangle$ & $\sigma(z)$ & $\langle$H$_\alpha\rangle$ & $\sigma($H$\alpha)$ & $\langle$[OII]$\rangle$ & $\sigma([$OII$])$ & $\langle$H$_{\delta}\rangle$ & $\sigma($H$\delta)$ & $\langle$H:K$\rangle$ & $\sigma($H:K$)$ & $\langle\sigma_*\rangle$ & $\sigma(\sigma_*)$ \\ 
      &  &  &  & \AA & \AA & \AA & \AA & \AA & \AA &  &  & kms$^{-1}$ & kms$^{-1}$ \\ 
    \hline 
    Parent & 8290 & 0.103 & 0.05 & 10.08 & 27.93 & 3.5 & 11.7 & -0.69 & 2.45 & 0.96 & 0.40 & 131.2 & 67.4 \\ 
    Photo & 3964 & 0.11 & 0.05 & -2.18 & 15.19 & -1.0 & 5.42 & 1.15 & 1.89 & 0.92 & 0.47 & 172.1 & 56.2 \\ 
    Spec-Photo & 2051 & 0.112 & 0.05 & -3.84 & 3.55 & -1.0 & 2.20 & 1.54 & 1.23 & 0.86 & 0.10 & 183.7 & 53.6 \\ 
    Bona fide & 231 & 0.16 & 0.06 & -3.48 & 2.69 & -1.59 & 1.97 & 1.95 & 1.10 & 0.84 & 0.10 & 272.8 & 34.6 \\ 
    \hline \\ 
    \end{tabular} 
\end{table*}

\begin{table*} 
    \centering 
    \caption{We report the median values and standard deviations for the following stellar population properties of each sample: stellar mass M$_*$, specific star formation rate sSFR, star formation rate SFR, metallicity Z/Z$_\odot$ and median mass-weighted age.} 
    \label{tab:med_vals2} 
    \begin{tabular}{c|cccccccccc} 
    \hline 
    Sample & $\langle$log(M$_*)\rangle$ & $\sigma($log(M$_*))$& $\langle$sSFR$\rangle$ & $\sigma($sSFR$)$ & $\langle$SFR$\rangle$ & $\sigma($SFR$)$ &$\langle$Z/Z$_{\odot}\rangle$ & $\sigma($Z/Z$_{\odot})$ & $\langle$age$\rangle$ & $\sigma($age$)$ \\ 
     & M$_{\odot}$ & M$_{\odot}$ & M$_{\odot}$yr$^{-1}$ & M$_{\odot}$yr$^{-1}$ & M$_{\odot}$yr$^{-1}$ & M$_{\odot}$yr$^{-1}$ & & & Gyr & Gyr \\ 
    \hline 
    Parent & 10.68 & 0.55 & -10.62 & 1.01 & -0.17 & 0.83 & 0.93 & 0.3 & 4.34 & 2.02  \\ 
    Photo & 10.92 & 0.42 & -11.77 & 0.65 & -0.93 & 0.69 & 0.94 & 0.26 & 6.19 & 1.44  \\ 
    Spec-Photo & 10.98 & 0.38 & -11.97 & 0.47 & -1.09 & 0.58 & 0.94 & 0.23 & 6.34 & 1.12  \\ 
    Bona Fide & 11.44 & 0.23 & -12.02 & 0.39 & -0.69 & 0.46 & 1.07 & 0.24 & 6.41 & 1.11  \\ 
    \hline 
    \end{tabular} 
\end{table*}

Table~\ref{tab:med_vals1} and Table~\ref{tab:med_vals2} show the median values and standard deviations for various properties of the parent, photometric, photo-spectro and bona fide CC samples. The median redshift of each subsequently selected sample increases from $z=0.103$ to $z=0.163$, with the largest jump between the spectroscopically selected galaxies and the stellar mass and stellar velocity dispersion selected samples. As the GAMA survey reports emission lines as positive values and absorption lines as negative values which means the median H$\alpha$ SNR and [OII] EW show a flip from emission to absorption between the parent sample and CC sample. In contrast, the median H$\delta$ EWs show a flip from absorption to emission, though negligible when considering the average uncertainty of H$\delta$ for the bona fide sample is $\sim0.74$~\AA and the lack of detectable H$\alpha$ and [OII] emission and low H:K ratio. Additionally, the same can be said of [OII] as the uncertainty for the bona fide sample is $\sim0.88$~\AA. It is interesting to note that the median H$\alpha$ SNR for the photo-spectro sample has the highest absorption out of all four samples. Fig.~\ref{fig:mcuts} shows the distribution of each of the samples for each spectral feature used in the selection criteria. In addition, the H:K ratio shows the least difference between the different samples, with all four median values remaining below 1. This suggests that even in the parent sample, the majority of galaxies are dominated by older stars, which is clear in Fig.~\ref{fig:mcuts}. It is obvious that the bona fide CC sample should have the highest median stellar velocity dispersions and stellar masses across the four groups, which is apparent in both Table~\ref{tab:med_vals2} and Fig.~\ref{fig:mcuts}, but properties also increase between the parent sample and the photo-spectro sample. 

\begin{figure*}
    \centering
    \includegraphics[width=\linewidth]{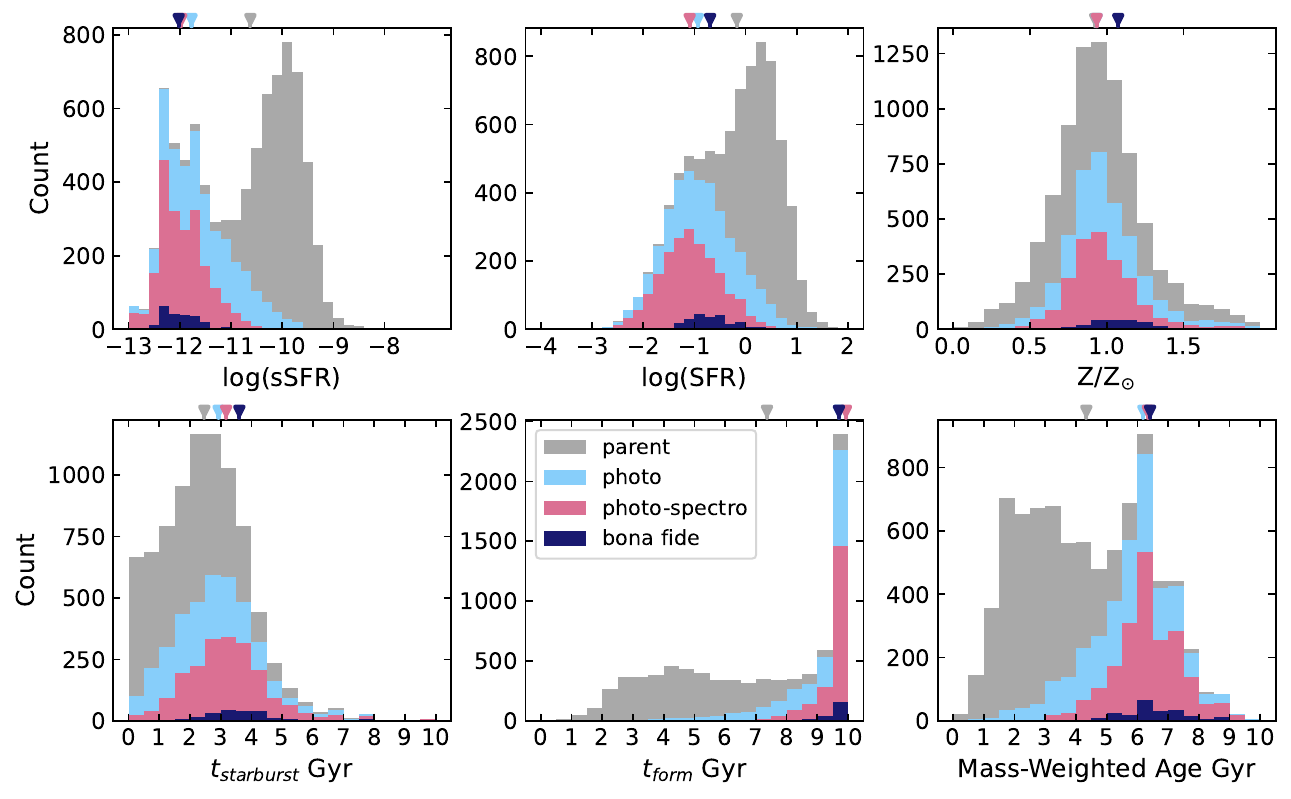}
    \caption[]{ The parent sample is shown in grey whilst the photo, photo-spectro and bona fide samples are shown in blue, pink and purple respectively. The median value for each sample is shown with an arrow. The upper panels, from left to right, show the sSFR, SFR and metallicity. The bottom panels, from left to right, show the median time since the last starburst ended $t_{\text{starburst}}$, the median age of the oldest stars in the galaxy $t_{\text{form}}$ and median mass-weighted age.}
    \label{fig:properties}
\end{figure*}

Finally, in Fig.~\ref{fig:properties} we show the distribution of galaxies in each selection sample according to their stellar population properties which are median values estimated by GAMA using the MagPhys SED fitting code. The sSFR shows that there is a clear bimodality within the parent sample as there are two peaks in the grey distribution at approximately $10^{-12}~\text{yr}^{-1}$ and $10^{-10}~\text{yr}^{-1}$. The peak at $10^{-12}~\text{yr}^{-1}$ contains the majority of the galaxies selected in the photo and photo-spectro subsamples and the entirety of the bona fide CC sample. The SFR and median mass-weighted age across the parent sample shows some bimodality with a parent sample peaks at approximately $-1~\text{M}_{\odot}\text{yr}^{-1}$ and $0.5~\text{M}_{\odot}\text{yr}^{-1}$ for SFR and $\sim2~\text{Gyr}$ and $6~\text{Gyr}$. There is less difference between the parent sample and subsequent subsample distributions across metallicity and median time since the last starburst ended ($\text{t}_{\text{starburst}}$), and the median values of each are higher for the bona fide CC sample by $\sim0.2$ and $\sim1.2~\text{Gyr}$ respectively. Finally, the median age of the oldest stars $\text{t}_{\text{form}}$ shows a fairly uniform distribution of the parent sample between $2-9~\text{Gyr}$ which has little overlap with the subsequent subsamples which have a strong peak at $\sim9.5~\text{Gyr}$ showing that the oldest stars of galaxies in the bona fide CC sample are significantly older than the majority of the parent sample. To summarise, we see that the properties of the CC candidates in the bona fide CC subsample indicate that these galaxies are old and quenched. This is especially clear when we consider the bimodality in the parent sample and the fact that the bona fide CC sample is found exclusively in the lower sSFR range, shown in Fig.~\ref{fig:nuvrj}

\subsection{Sample Binning} \label{binning}
To calculate the Hubble parameter using the differential age-redshift relation, we first need to split the cosmic chronometer sample into high and low stellar velocity dispersion $\sigma_*$ regimes. To do this, we simply take the median value $\sigma_*=272.77~\text{kms}^{-1}$ and anything above this is considered to be in the ``high''-$\sigma_*$ sample whereas anything below this is assigned to the ``low''-$\sigma_*$ sample. Our total number of CC candidates is 231 so this results in a ``high''-$\sigma_*$ group with 116 galaxies and a ``low''-$\sigma_*$ group with 115 galaxies between $0.0384\lesssim  z \lesssim 0.335$. 

Additionally, to calculate $H(z)$ we must bin these two groups based on redshift as the width of each redshift bin must correspond to a cosmic time that is larger than the average age uncertainty to overcome the statistical scatter associated with the ages. Our average age uncertainty is calculated using the 16th and 84th percentiles for the median mass-weighted ages from the GAMA survey as the difference between these values corresponds to $1\sigma$. We find $\Delta\text{age}\simeq1.5~\text{Gyr}$ which corresponds to a redshift interval of $\Delta z \simeq 0.103$ which allows us to create three bins based on redshift between $0.03 \leq z \leq 0.339$. In comparison to previous CC studies, $\Delta z \simeq 0.103$ are wide redshift bins which is not ideal. Though the bins need to be large enough to overcome statistical scatter from the ages of the CCs they also need to remain small enough that there is minimal age evolution within each bin.

\begin{figure}
    \centering
    \includegraphics[width=\linewidth]{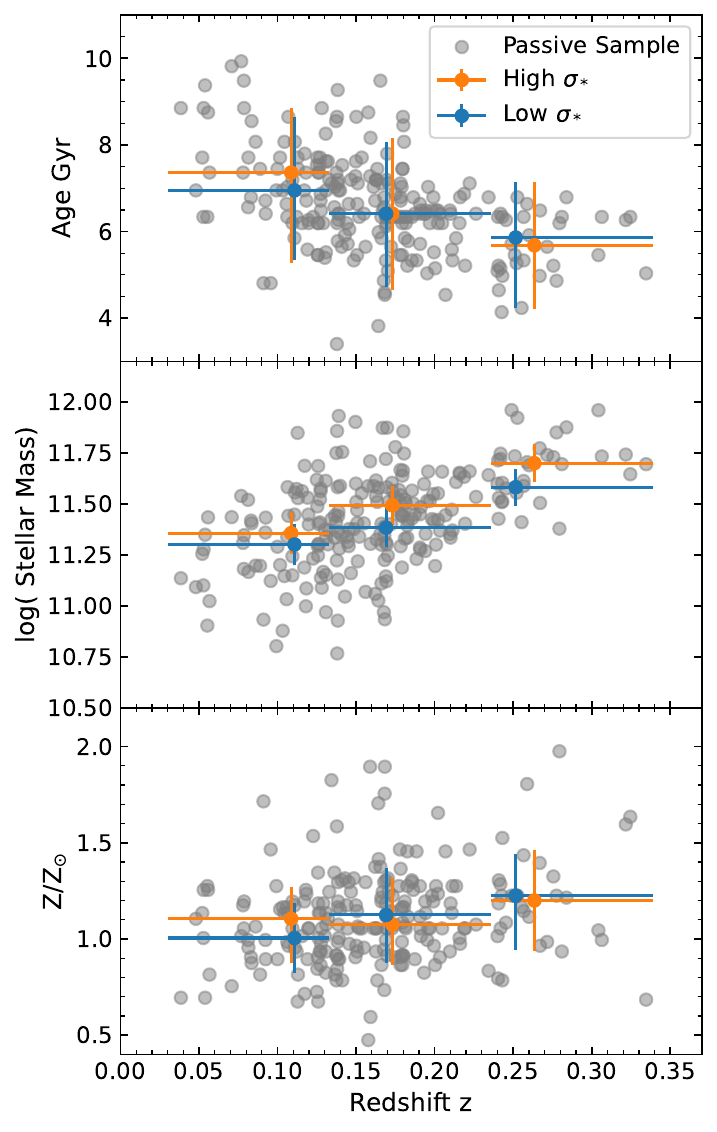}
    \caption{We compare the median mass-weighted age (top panel), stellar mass (middle panel) and metallicity (bottom panel) across redshift for the bona fide CCs by splitting the CC subsample into high and low stellar velocity dispersion groups, shown in orange and blue respectively, we then bin them based on their redshift. The width of the bins is shown with horizontal error bars and the median value for each bin is shown with blue or orange points. The vertical error bars correspond to age, stellar mass and metallicity uncertainties. The grey points represent the total bona fide passive CC sample. }
    \label{fig:binned}
\end{figure}

We show the result of this binning process in Table~\ref{tab:binned} along with the median values of various properties for each bin. For each redshift bin in both $\sigma_*-$regimes we find that $\sigma_*$, stellar mass and metallicity increase with increasing redshift whilst median mass-weighted age decreases with increasing redshift which can also be seen in Fig.~\ref{fig:binned}. The increase of stellar mass with redshift indicates that the parent sample is not statistically complete. Typically, CCs should have a fairly uniform stellar mass across redshift, however if they do not it becomes difficult to disentangle the ages of the galaxies and therefore their evolution history from other effects relating to stellar mass. This is demonstrated in the CC sample assembled by \cite{2023ApJS..265...48J}, in which their sample has stellar masses between $10^{10}-10^{12}~\text{M}_{\odot}$ across a redshift range between $0.6<z<1$. 

\begin{table*}
    \centering
    \caption{The number of galaxies in each $\sigma_*$ regime are binned across redshift. For each bin, we show median values for redshift, $\sigma_*$, stellar mass, metallicity and median mass-weighted age.}
    \begin{tabular}{c|cccccc}
        Bin & Count & $\langle z \rangle$ & $\langle \sigma_* \rangle$~kms$^{-1}$ & $\langle \text{log(M}_*) \rangle$ & $\langle Z \rangle ~Z_\odot$ & $\langle \text{age} \rangle$~Gyr\\
        \hline 
        High-$\sigma_*$ & 116 & & & & & \\
        \hline
        Bin 1 & 26 & 0.11 & 288.8 & 11.4 & 1.12 & 7.36 \\
        Bin 2 & 70 & 0.17 & 291.8 & 11.5 & 1.08 & 6.41 \\ 
        Bin 3 & 20 & 0.26 & 307.7 & 11.7 & 1.20 & 5.68 \\
        \hline 
        Low-$\sigma_*$ & 115 & & & & & \\
        \hline
        Bin 1 & 45 & 0.11 & 257.4 & 11.3 & 1.01 & 6.95 \\
        Bin 2 & 58 & 0.17 & 259.5 & 11.4 & 1.13 & 6.41 \\ 
        Bin 3 & 12 & 0.25 & 260.6 & 11.6 & 1.23 & 5.86 \\
        \hline 
    \end{tabular}
    \label{tab:binned}
\end{table*}

\begin{figure}
    \centering
    \includegraphics[width=\linewidth]{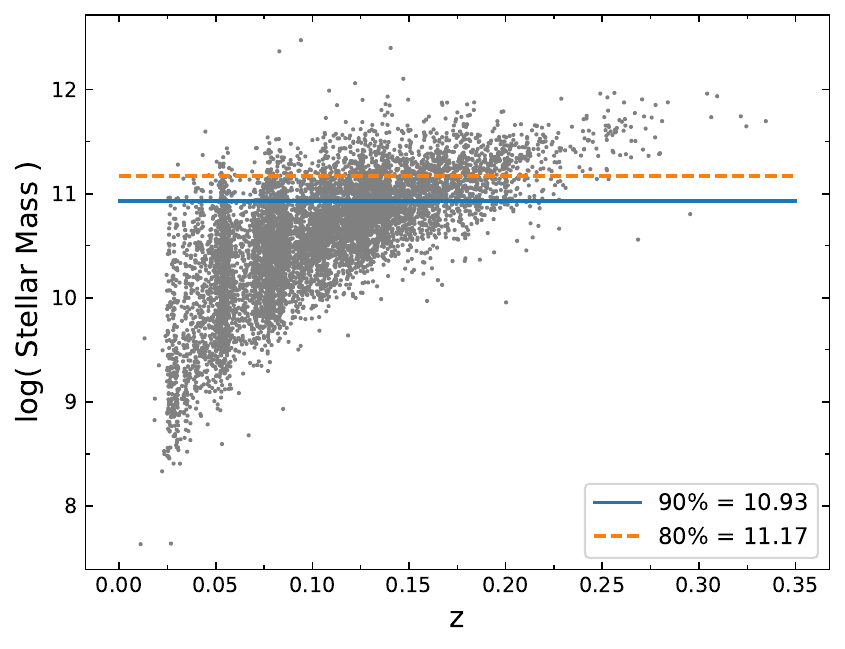}
    \caption{Here we show the redshift and stellar mass of the entire sample of 8290 galaxies to assess the lower limit of mass detected by GAMA. We mark the 80\% and 90\% completeness levels of the sample with a dashed orange line and solid blue line respectively. The stellar mass limit needed for an 80\% and 90\% completeness is $\sim10^{11.2}~\text{M}_{\odot}$ and $\sim10^{10.9}~\text{M}_{\odot}$ respectively which corresponds to a redshift limit of $z\simeq0.25$ or $z\simeq0.18$.}
    \label{fig:mass_z}
\end{figure}

\begin{figure*}
    \centering
    \includegraphics[width=\linewidth]{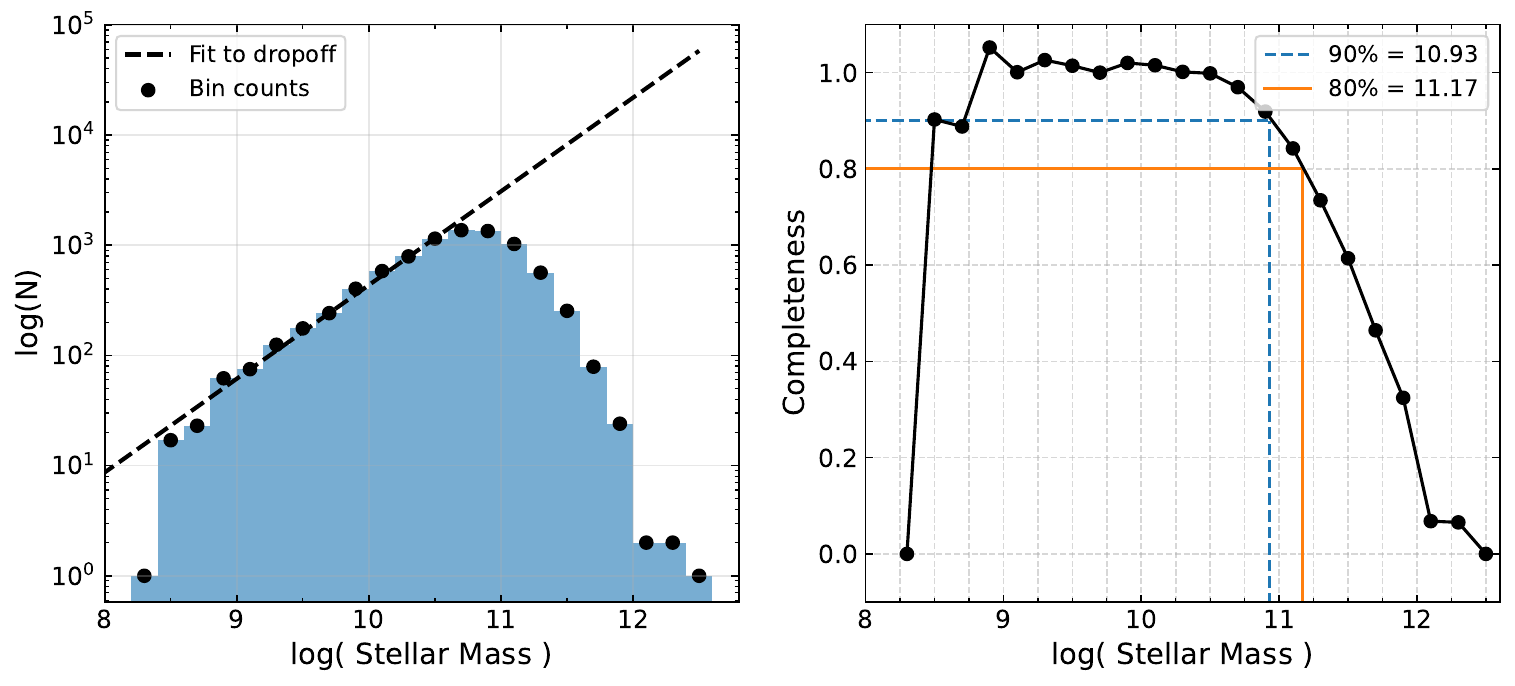}
    \caption{The \textit{left panel} shows a logged histogram of the stellar mass of the entire parent sample in which we fit a dashed line of best to the region where mass is increasing between $8\lesssim\text{log(M}_{\odot}/\text{M}_*)\lesssim10.8$. We calculate the ratio between this line and the bins of mass to create the plot of completeness, shown in the \textit{right panel}. We indicate the points at which the mass is 80\% and 90\% complete with a solid orange line and dashed blue line respectively which are at approximately $\sim10^{11.2}$~M$_{\odot}$ and $\sim10^{10.9}$~M$_{\odot}$.}
    \label{fig:completeness}
\end{figure*}

To investigate the completeness of our sample in relation to stellar mass, we perform a completeness test to assess at what redshift our sample becomes incomplete in terms of stellar mass. We show the stellar masses for the total sample across redshift in Fig.~\ref{fig:mass_z} along with the completeness limits calculated using Fig.~\ref{fig:completeness}. To find these limits we bin stellar mass for the entire parent sample and find the point at which the completeness drops off by calculating the ratio between each bin and the linear fit up to the drop-off, shown in Fig.~\ref{fig:completeness}. Using this method, we find the sample is 80\% complete up to $z\simeq0.25$ and 90\% complete up to $z\simeq0.18$. As our CC sample is concentrated towards the higher-redshift end of the parent sample, this may explain the increase in median redshift of $\Delta z\simeq0.05$. Since the parent sample becomes only 80\% complete in stellar mass beyond $z\simeq0.25$, the relative fraction of lower-mass galaxies is likely to decrease towards the upper end of the redshift range. Consequently, selecting only high-mass galaxies may preferentially retain galaxies at higher redshifts, resulting in a shift towards a higher median redshift.

\section{Dependence of Statistical Uncertainty Contribution on Sample} \label{stat_unc}
The choice of selection criteria for identifying cosmic chronometers varies across the literature, with some criteria like photometric selection and [OII] emission being more widely used and other criteria such as H$\delta$ absorption and the H:K ratio. The benefit of using less strict selection criteria increases the overall CC sample size which can help to reduce the statistical uncertainty on the resulting $H_{0}$ or $H(z)$ values but can also introduce more uncertainty due to the impurity of the sample. As such, we briefly investigate the contribution of statistical uncertainty due to the impurity for our sample here. To do this, we calculate a baseline value for $H(z)$ using the CC sample chosen with the strictest selection criteria. We then calculate different values for $H(z)$ using CC samples chosen with less strict criteria in order to evaluate the leading contribution of impurity uncertainty for our sample. Please note, this is only an evaluation of uncertainty relating to the purity of the selected CC sample and a more in-depth description of the H(z) calculation and uncertainty determination (including both systematic and statistical uncertainty) is in our follow-up paper. To isolate the impact of the galaxy selection criteria, only statistical uncertainties are shown in this analysis. Systematic uncertainties associated with the stellar population synthesis modelling are common to all selection tests and therefore do not affect the relative comparison between different selection strategies.

\subsection{Baseline H(z)}
We calculate a baseline $H(z)$ value with the bona fide CC sample by using the differential age-redshift relation: 

\begin{equation}
    H(z_{\text{eff}}) \equiv \frac{-1}{1+z_{\text{eff}}}\frac{\Delta z}{\Delta t}
\end{equation}

where $z_{\text{eff}}$ is the effective redshift between two populations of CCs separated by a redshift $\Delta$ and difference in age $\Delta t$ \citep{2002ApJ...573...37J}. We calculate $H(z)$ for both the high- and low-$\sigma_*$ samples between bins 1 and 3 for an effective redshift of $z_{\text{eff}} = (z_1 +z_3)/2$ as this ensures the evolution of age over the difference in redshift $\Delta z$ is greater than the scatter associated with the age uncertainties. We report the total number of galaxies in bin 1 and bin 3 for each regime alongside the $\Delta z$, $\Delta \text{age}$ and $z_{\text{eff}}$ between bins 1 and 3 in Table~\ref{tab:hz}. 

We use a bootstrapping resampling method to randomly sample from bin 1 and bin 3 in order to calculate $H(z)$ and its associated statistical uncertainties from the resulting percentile distributions. We perform 100,000 bootstrapping iterations, in which each iteration resamples both bin 1 and bin 3 simultaneously. We then calculate $\Delta z$, $\Delta t$, $z_{\text{eff}}$ and finally $H(z)$ for each iteration, shown in Table~\ref{tab:hz}.

We combine the low- and high-velocity dispersion measurements using inverse-variance weighting, with the mean of the upper and lower bootstrap uncertainties adopted as an effective symmetric uncertainty for the weighting. The statistical uncertainty of the combined measurement is then obtained by propagating the asymmetric uncertainties separately. We find the combine, error-weighted baseline to be $H(z=0.183)= 85.54 ^{+ 16.21 }_{- 32.93 }$~kms$^{-1}$Mpc$^{-1}$, as shown in Table~\ref{tab:hz}. 

\begingroup
\renewcommand{\arraystretch}{1.3}
\begin{table*}
    \centering
    \begin{tabular}{l|ccccccc}
        Test & Total No. Passed & No. in Bin 1+3 & $\Delta z$ & $\Delta$~age & $z_{\text{eff}}$ & $H(z_{\text{eff}})$ & $\Delta H(z)$ \\
         & & & & Gyr & & kms$^{-1}$Mpc$^{-1}$ & kms$^{-1}$Mpc$^{-1}$ \\
        \hline
        \hline 
        High $\sigma_*$ \\
        \hline 
        Only NUVrJ & 170 & 66 & 0.153 & -1.712 & 0.184 & $ 72.86 ^{+ 12.91 }_{- 16.27 }$ & \\
        Only Spec & 133 & 53 & 0.148 & -1.712 & 0.186 & $ 74.19 ^{+ 14.6 }_{- 29.13 }$ & \\
        Only High Mass & 27 & 15 & 0.157 & -1.374 & 0.202 & $ 96.99 ^{+ 36.2 }_{- 132.49 }$ & \\
        Only Low Mass & 175 & 69 & 0.148 & -1.512 & 0.185 & $ 76.19 ^{+ 13.93 }_{- 22.08 }$ & \\ 
        Bona fide & 116 & 46 & 0.155 & -1.68 & 0.186 & $ 76.89 ^{+ 12.88 }_{- 16.26 }$ & \\
        \hline 
        Low $\sigma_*$ \\
        \hline 
        Only NUVrJ & 169 & 74 & 0.147 & -1.55 & 0.179 & $ 76.89 ^{+ 12.88 }_{- 16.26 }$ & \\
        Only Spec & 132 & 64 & 0.141 & -1.09 & 0.181 & $ 99.01 ^{+ 33.79 }_{- 82.6 }$ & \\
        Only High Mass & 27 & 9 & 0.169 & -1.323 & 0.177 & $ 87.86 ^{+ 40.69 }_{- 78.95 }$ & \\
        Only Low Mass & 174 & 74 & 0.147 & -1.617 & 0.18 & $ 72.99 ^{+ 12.11 }_{- 13.11 }$ & \\ 
        Bona fide & 115 & 57 & 0.142 & -1.09 & 0.18 & $ 98.71 ^{+ 33.62 }_{- 84.38 }$ & \\
        \hline 
        Error-weighted \\
        \hline 
        Only NUVrJ & 339 &  &  &  & 0.181 & $ 74.88 ^{+ 9.12 }_{- 11.5 }$ & 10.67 \\
        Only Spec & 265 &  &  &  & 0.183 & $ 77.26 ^{+ 13.46 }_{- 27.5 }$ & 8.28 \\
        Only High Mass & 54 &  &  &  & 0.191 & $ 90.91 ^{+ 29.66 }_{- 68.74 }$ & -5.37 \\
        Only Low Mass & 349 &  &  &  & 0.182 & $ 74.04 ^{+ 9.33 }_{- 11.41 }$ & 11.50 \\
        Bona fide & 231 &  &  &  & 0.183 & $ 85.54 ^{+ 16.21 }_{- 32.93 }$ &  \\
        \hline
    \end{tabular}
    \caption{We report the total number of galaxies in bins 1 and 3, their difference in redshift $\Delta z$, difference in age $\Delta \text{age}$, effective redshift and calculated $H(z)$ value with statistical uncertainties found using a bootstrapping method for the high- and low-$\sigma_*$ regimes. We also report the error-weighted results along with the total number of galaxies selected with each different set of selection criteria. Note that the NUVrJ-only and spec-only samples still incorporate lower mass and velocity dispersion prescriptions into their selection criteria.}
    \label{tab:hz}
\end{table*}

\subsection{Purity of Sample and Statistical Uncertainty Contribution}
To estimate the contribution of statistical uncertainty associated with the selection of cosmic chronometers, we compare the baseline $H(z)$ with $H(z)$ calculated using different samples of potential CCs. The first selection is a photometric-based one that uses NUVrJ colours, stellar mass, and stellar velocity dispersion; the second uses different spectroscopic tracers together with stellar mass and velocity dispersion; and the third uses only stellar mass and velocity dispersion but varies between the more strict and less strict prescriptions according to \cite{2018ApJ...868...84M}. Stellar mass and velocity dispersion are used in all three selections to ensure that the resulting samples remain homogeneous enough for a suitable $H(z)$ analysis across the full redshift range. In particular, the median velocity dispersion is required to be representative of both the low- and high-redshift regimes. When velocity dispersion is not included as a selection criterion, the resulting samples have systematically higher median velocity dispersions, which leads to there being no galaxies satisfying the selection in the highest-redshift bin. Thus, keeping stellar mass and velocity dispersion fixed allows the different selections to probe the effect of varying the additional selection criteria, while avoiding changes that would make the high redshift measurement unconstrained. The variation in the resulting $\Delta H(z)$ measurements between these selection schemes is therefore used as an estimate of the statistical uncertainty associated with the choice of cosmic chronometer sample where $\Delta H(z)=H(z)_{\text{baseline}}-H(z)_{\text{subsample}}$.
 
Firstly, Table~\ref{tab:hz} shows the results for the high and low regimes alongside the error-weighted $H(z)$ value when we use the NUVrJ diagram to select passive galaxies and further restrict the sample with log$(M_{*}/M_{\odot}$)~>~10.75 and $\sigma_{*}$~>~250~kms$^{-1}$ and find $H(z=0.181)=74.88 ^{+ 9.12 }_{- 11.5 }$~kms$^{-1}$Mpc$^{-1}$ with a difference from the baseline equal to $\Delta H(z)=10.67$~kms$^{-1}$Mpc$^{-1}$. This is a similar value to the $H(z)$ calculated using the subsample defined by only spectroscopic, mass and $\sigma_*$ limits, in which we find $H(z=0.183)=77.26 ^{+ 13.46 }_{- 27.5 }$~kms$^{-1}$Mpc$^{-1}$ with a difference from the baseline equal to $\Delta H(z)=8.28$~kms$^{-1}$Mpc$^{-1}$. Additionally, the photometrically selected subsample and the low-mass only subsample result in the smallest statistical uncertainties as these are the largest of all the subsamples, with 339 and 349 galaxies in each sample respectively.

For the high-mass, high-velocity dispersion subsample we use galaxies with log$(M_{*}/M_{\odot}$)~>~11 and $\sigma_{*}$~>~300~kms$^{-1}$. For this subsample we find $H(z=0.191)=90.91 ^{+ 29.66 }_{- 68.74}$~kms$^{-1}$Mpc$^{-1}$ and $\Delta H(z)=-5.37$~kms$^{-1}$Mpc$^{-1}$ which is significantly different to the lower mass and stellar velocity dispersion subsamples in which $H(z=0.182)=74.04 ^{+ 9.33 }_{- 11.41}$ with a difference from the baseline equal to $\Delta H(z)=11.5$. This difference is likely due to the higher-mass prescription yielding only 54 galaxies in the final subsample whereas the low-mass subsample contains 349 galaxies. This difference obviously significantly increases the statistical uncertainty of $H(z)$.

Finally, we compare the difference in $H(z)$ when only specific spectroscopic selection criteria are used in addition to the lower mass and velocity dispersion limits in Table~\ref{tab:spec_hz}. We find the the largest difference occurs when on only [OII] emission is used in which $H(z=0.182)=69.49 ^{+ 9.22 }_{- 15.53 }$~kms$^{-1}$Mpc$^{-1}$ and $\Delta H(z)=16.05$~kms$^{-1}$Mpc$^{-1}$ whereas the H:K ratio subsamples has the smallest difference with $H(z=0.183)=77.28 ^{+ 13.49 }_{- 27.55 }$~kms$^{-1}$Mpc$^{-1}$ and $\Delta H(z)= 8.26$~kms$^{-1}$Mpc$^{-1}$. 

\begin{table*}
    \centering
    \begin{tabular}{l|ccccccc}
        Test & Total No. Passed & No. in Bin 1+3 & $\Delta z$ & $\Delta$~age & $z_{\text{eff}}$ & $H(z_{\text{eff}})$ & $\Delta H(z)$ \\
         & & & & Gyr & & kms$^{-1}$Mpc$^{-1}$ & kms$^{-1}$Mpc$^{-1}$ \\
        \hline
        \hline 
        High $\sigma_*$ \\
        \hline 
        Only [OII] &  159 & 63 &  0.154 & -1.809 & 0.184 & $ 68.86 ^{+ 10.54 }_{- 15.96 }$ &  \\
        Only H$\alpha$ & 157 & 61 & 0.153 & -1.647 & 0.184 & $ 78.0 ^{+ 13.52 }_{- 25.63 }$ &  \\
        Only H$\delta$ & 173 & 68 & 0.149 & -1.545 & 0.185 & $ 74.99 ^{+ 13.77 }_{- 21.25 }$ &  \\
        Only H:K & 133 & 53 & 0.148 & -1.713 & 0.186 & $ 74.19 ^{+ 14.64 }_{- 29.21 }$ &  \\
        \hline 
        Low $\sigma_*$ \\
        \hline 
        Only [OII] & 159 & 67 & 0.144 & -1.615 & 0.180 & $ 73.23 ^{+ 13.39 }_{- 51.14 }$ &  \\
        Only H$\alpha$ & 157 & 64 & 0.142 & -1.677 & 0.181 & $ 70.2 ^{+ 12.42 }_{- 25.69 }$ &  \\
        Only H$\delta$ & 172 & 73 & 0.147 & -1.555 & 0.179 & $ 75.99 ^{+ 13.81 }_{- 17.25 }$ &  \\
        Only H:K & 132 & 64 & 0.141 & -1.088 & 0.181 & $ 98.88 ^{+ 33.7 }_{- 82.33 }$ &  \\
        \hline 
        Error-weighted \\
        \hline 
        Only [OII] & 318 &  &  &  & 0.182 & $ 69.49 ^{+ 9.22 }_{- 15.53 }$ & 16.05 \\
        Only H$\alpha$ & 314 &  &  &  & 0.183 & $ 73.99 ^{+ 9.16 }_{- 18.15 }$ & 11.55 \\
        Only H$\delta$ & 345 &  &  &  & 0.182 & $ 75.55 ^{+ 9.83 }_{- 13.44 }$ & 10.0 \\
        Only H:K & 265 &  &  &  & 0.183 & $ 77.28 ^{+ 13.49 }_{- 27.55 }$ & 8.26 \\
        \hline
    \end{tabular}
    \caption{Here we report the total number of galaxies in bins 1 and 3, their difference in redshift $\Delta z$, difference in age $\Delta \text{age}$, effective redshift and calculated $H(z)$ value with statistical uncertainties found using a bootstrapping method for the high- and low-$\sigma_*$ regimes for each subsample. We also report the error-weighted results along with the total number of galaxies selected with each different set of selection criteria. All subsamples still incorporate lower mass and velocity dispersion prescriptions into their selection criteria.}
    \label{tab:spec_hz}
\end{table*}
\endgroup

\begin{figure}
    \centering
    \includegraphics[width=\linewidth]{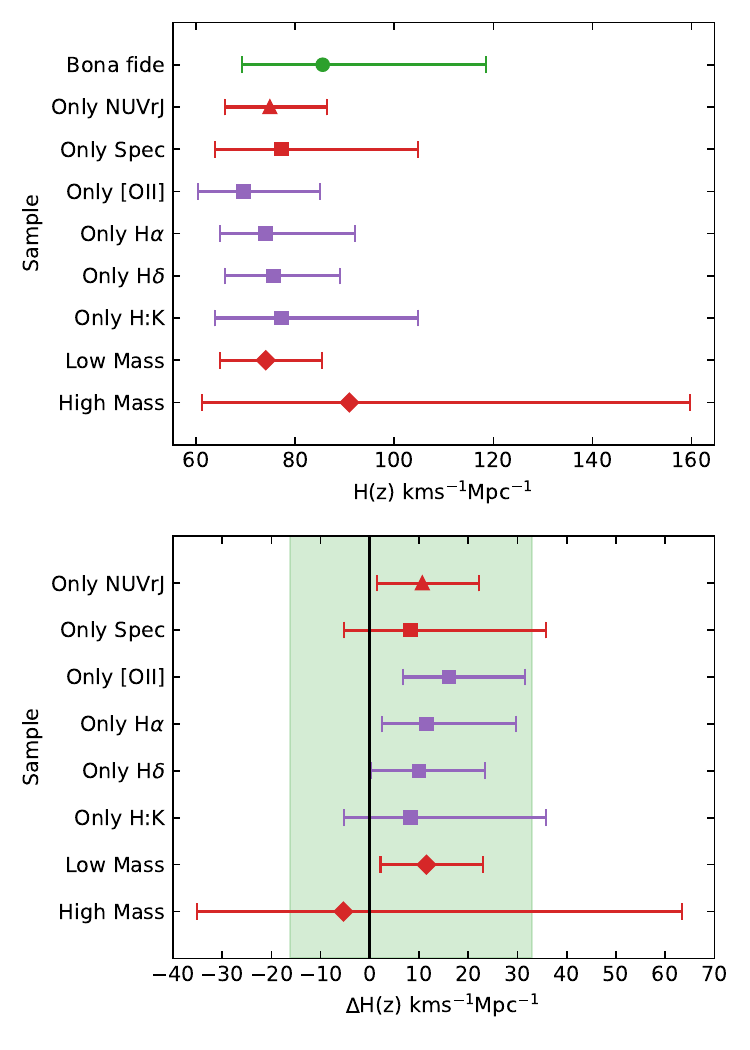}
    \caption{In the top panel we show the calculated $H(z)$ and statistical uncertainties for each subsample including the baseline value calculated with the bona fide CC sample. In the lower panel we show the relative difference in $H(z)$ with respect to the baseline which is marked at 0. The statistical uncertainty of the baseline is shown with a shaded green rectangle.}
    \label{fig:diff_hz}
\end{figure}

Fig.~\ref{fig:diff_hz} shows the $H(z)$ values for all samples and the relative difference from the baseline. This shows clearly which selection criteria have the most effect on the statistical uncertainty of $H(z)$. The photometric cut and the low-mass only cut have the lowest statistical uncertainties as they yield the largest number of galaxies in each sample.

\section{Discussion} \label{disc}
The GAMA survey \citep{2009A&G....50e..12D, 2015MNRAS.452.2087L} has spectroscopic and photometric observations for $\sim$300,000 galaxies with a spectral resolution of $R\approx1300$ over range that is spliced at 5700~{\AA} from a blue spectrum covering $3750\leq \lambda \leq 5850$~{\AA} and a red spectrum across $5650\leq \lambda \leq 8850$~{\AA} to yield full spectra across $3750\leq \lambda \leq 8850$~{\AA} \citep{2017MNRAS.465.2671G}. After cleaning and cross-matching with SDSS catalogue data we yield a sample of 8290 galaxies with redshifts between $0.003<z<0.42$ to apply \cite{2018ApJ...868...84M} selection cuts to. Out of 8290 total galaxies, we identify 231 CCs with only 199 of which having matching SDSS photometric images available. This is approximately 2.8\% of the clean sample and approximately 0.08\% of the total survey. In comparison, the Large Early Galaxy Astrophysics Census (LEGA-C) survey \citep{2018ApJS..239...27S, 2016ApJS..223...29V} spans $\sim4000$ galaxies between $0.6<z<1$ in the COSMOS field. The LEGA-C flux-calibrated spectra have a spectral resolution of $R\sim3500$ with a wavelength range between $6300\leq \lambda \leq 8800$~{\AA}. \cite{2022ApJ...927..164B} identifies 350 CCs from 1622 sources after cross-matching with the COSMOS2015 catalogue \citep{2016ApJS..224...24L}. This is a 9\% yield from the entire LEGA-C survey and an 22\% yield for the cross-matched final sample. This sample is selected with fewer selection cuts than we use however they provide a full analysis of the spectral indices and H:K ratio. They make three cuts using the same NUVrJ selection as us, the same cut for for [OII] emission at EW[OII]~$>5$~{\AA} and a visual inspection of the spectra that pass the first two cuts which yields 350 CCs. The visual inspection removes an galaxies with strong [OII] $\lambda 3727$ and [OIII] $\lambda 5007$ emission lines. However, their final CC sample is limited once full spectral fitting is completed for the age estimations, the final sample ends up being 140 CCs with corresponding ages, metallicities and SFHs.

It is expected that more CCs are found at higher redshifts and by no means is this an extensive investigation but we yield about four times less at a much lower redshift. However, \cite{2023A&A...679A..96T} identifies 49 CCs from 2087 galaxies across $1 < z < 1.5$ in the fourth data release of the VANDELS survey \citep{2018MNRAS.479...25M, 2020MNRAS.492.2128P, 2021A&A...647A.150G} which is approximately a 2.3\% yield. The VANDELS survey has a mean spectral resolution of $R\simeq650$ across an average wavelength range of $4800\leq \lambda \leq 9800$~{\AA}. This sample is made using five selection cuts, two of which relate to the chosen redshift range as their method relies on the differential age-D4000 relation rather than the age-redshift relation. Additionally, they use a UVJ selection cut rather than an NUVrJ cut in which $U-V>0.88(V-J)+0.49$, $U-V>1.2$ and $V-J>1.6$. They use the same [OII] emission cut at EW[OII]$>5$~{\AA} and the \cite{1984AJ.....89.1238R} H:K ratio in which they select galaxies with H:K$<1.3$. Finally, they visually inspect all spectra for residual emission lines or calibration issues. 

\cite{2016JCAP...05..014M} yields $\sim$130,000 CCs identified from the approximately 1.5 million galaxies observed in the DR9 Baryon Oscillation Spectroscopic Survey (BOSS) \citep{2005ApJ...633..560E, 2013AJ....145...10D} between $0.3<z<0.5$. This is a yield of 8.7\% of the total BOSS data and 15\% out of the 848,697 galaxies on which they perform their selection cuts on. BOSS spectra have a resolution of $R\sim2000$ across an observed wavelength range of $3750\leq \lambda \leq 10000$~{\AA}. This work uses three selection cuts, the first of which considers galaxies with $g-i>2.35$ to be passive. Additionally, they cut for [OII] emission EW[OII]~$>5$~{\AA} alongside a signal-to-noise ratio cut at SN(EW)~$<2$ and H$\alpha$, [OIII] $\lambda 5008$ and H$\beta$ at SN(EW)~$>2$. Finally, they include a D4000 cut at SN(D4000)~$>10$ to ensure low-quality measurements are not included when they calculate the differential age-D4000 relation.

It is important to bear in mind that although the CC sample is chosen with strict selection cuts in order to minimise the presence of young stellar populations, ensuring these galaxies are amongst the oldest objects at their respective redshifts, there may be contamination as a result of issues with the spectroscopy such as aperture effects. For example, GAMA uses AAOmega spectroscopy using the Anglo-Australian Telescope \citep{2004SPIE.5492..410S} which has an aperture diameter of 2~arcseconds which means low redshift galaxies may not be fully covered by the fibre therefore signatures from younger populations may be missed. Similarly, the BOSS spectrograph has a circular fibre of 2~arcseconds however VANDELS and LEGA-C are both observed using the VIsible Multi-Object Spectrograph (VIMOS) that has a rectangular slit with a fixed width of 1~arcsecond. VANDELS is configured for deep, faint, high-redshift galaxies and can cover the entirety of galaxies between $1 < z < 1.5$ whereas LEGA-C will only cover the inner regions of galaxies between $0.6 < z < 1$. This means LEGA-C galaxies could have younger stellar populations that are not covered, therefore leading to more contamination of the final CC sample than in the VANDELS CC sample. Our CC candidates which are observed with GAMA are the most affected by these aperture effects as they have relatively low redshifts with a small fibre diameter which means larger portions of the outer regions of galaxies may be missed in comparison to other surveys. Additionally, it is possible that if there is a significant amount of mass from younger stars outside of the aperture, then the mass-weighted ages may be skewed slightly higher which means caution may need to be used when comparing our CC sample to samples in other surveys. To avoid the inclusion of any systematic bias when using our sample with other samples we suggest deriving ages using the same method and rebinning based on the new age uncertainties.

Finally, all of the above mentioned studies that calculate the H:K ratio use the \cite{1984AJ.....89.1238R} method of calculating the H:K ratio which takes the direct pixel values of the troughs for the Ca II H and K lines however we use the \cite{2019MNRAS.490..455P} method which accounts for the significance of the continuum line. With higher resolution spectroscopy could potentially resolve the H$\epsilon$ in the dip due to backfilling. This would give us an indication of \emph{both} the Ca and H lines simultaneously rather than them being blurred together. When considering all the possible selection cuts in comparison to our chosen cuts, we find that other work that relies on fewer cuts find higher numbers of CCs. 

\section{Summary} 
In this work, we present 231 massive, passively evolving elliptical galaxies that are good candidates for cosmic chronometry from the GAMA survey at $0.03<z<0.4$ which can be used in addition to other samples of CCs for calculations of the Hubble constant. We use well-tested traditional methods of classification to ensure the final CC sample is free from contamination of younger populations. In addition, when we compare the bona fide CC candidates with galaxies that do not pass the selection criteria we find:
\begin{enumerate}
    \item The median redshift for the bona fide CC subsample is $\sim0.05$ higher than that of the parent, photo and photo-spectro subsamples.
    \item The bimodality of bluer, star-forming galaxies and redder, passive galaxies is clearly shown in our parent sample. The entirety of the bona fide CC subsample falls within the redder region of the two peaks, showing the subsample is likely to be passive. 
    \item The bona fide CC sample shows higher amounts of H$\alpha$ and [OII] absorption than the parent sample and higher H$\delta$ emission. Though the amount of absorption and emission would still be considered negligible with regard to the uncertainty for each corresponding EW.
    \item H:K values for all subsamples indicate a dominance of an older stellar population, with the bona fide CC subsample having the lowest H:K ratio with H:K~$\simeq0.84$ 
    \item The median sSFR for the bona fide CC subsample is $10^{-12}~\text{Myr}^{-1}$ whereas the median sSFR for the parent sample is $10^{-10}~\text{Myr}^{-1}$. This shows the bona fide CC sample is much more quiescent which can also be seen in the bimodality of sSFR in Fig.~\ref{fig:properties}.
    \item The bona fide CC subsample has slightly higher metallicity in comparison to the parent with $Z/Z_\odot=1.08$ and $Z/Z_\odot=0.925$ respectively.
    \item The median mass-weighted ages of the bona fide CC subsample are typically much older than the parent subsample, with median values of 6.41~Gyr and 4.34~Gyr respectively. Additionally, the parent sample shows a bimodality across the median mass-weighted ages, in which the bona fide CC subsample falls entirely in the older peak.
\end{enumerate}

The number of cosmic chronometers that we identify in the GAMA survey is mostly limited due to the quality of data relating to the information we need in order to classify them using traditional criteria. To highlight this issue, out of approximately $\sim300,000$ galaxies observed in the GAMA survey, approximately $\sim9,000$ have everything we need to classify them. On top of this, the CC sample is only statistically complete up to $z\sim0.2$ when considering stellar mass which may be an issue when calculating $H(z)$ or $H_0$ unless this gap is filled. We find the sample is 80\% complete up to a redshift of $z\simeq0.25$ and 90\% complete up to $z\simeq0.18$. With this in mind, we hope to utilise machine learning with our sample of CCs across different large scale surveys such as SDSS, Euclid and Rubin LSST to identify cosmic chronometers without relying on spectroscopic data as the initial first step of identification to dramatically increase the pool of promising cosmic chronometer candidates. Therefore, we plan to investigate the use of machine learning to identify cosmic chronometers in the follow-up to this paper which utilises the CC sample assembled here.

\section{Acknowledgements}
We would like to acknowledge the Viper High Performance Computing facility of the University of Hull and its support team. Additionally, the GAMA survey which is a joint European-Australasian project based around a spectroscopic campaign using the Anglo-Australian Telescope. The GAMA input catalogue is based on data taken from the Sloan Digital Sky Survey and the UKIRT Infrared Deep Sky Survey. Complementary imaging of the GAMA regions was obtained by a number of independent survey programmes including GALEX MIS, VST KiDS, VISTA VIKING, WISE, Herschel-ATLAS, GMRT and ASKAP providing UV to radio coverage. GAMA is funded by the STFC (UK), the ARC (Australia), the AAO, and the participating institutions. The GAMA website is https://www.gama-survey.org/ .

\section{Data Availability}
Our cosmic chronometer sample will be made available through SIMBAD upon publication. 



\bibliographystyle{mnras}
\bibliography{bibliography}




\bsp	
\label{lastpage}
\end{document}